\documentclass[]{aa}  
\usepackage{graphicx}
\usepackage{txfonts}
\usepackage{natbib}
\begin{document}
   \title{Long-term lightcurves from combined unified very high energy $\gamma$-ray data}

   \author{M. Tluczykont \inst{1,2}
          \and E. Bernardini \inst{1}
 	  \and K. Satalecka \inst{1}
          \and R. Clavero \inst{1,3}
	  \and M. Shayduk \inst{1,4}
	  \and O. Kalekin \inst{1,5}
%          C. Ptolemy\inst{2}\fnmsep\thanks{Just to show the usage
%          of the elements in the author field}
          }

   \institute{DESY
              Platanenallee 6, 15738 Zeuthen, Germany\\
              \email{martin.tluczykont@desy.de}
         \and
             Now at Universit\"at Hamburg, Luruper Chaussee 149, 22761 Hamburg, Germany
         \and
             Now at Isaac Newton Group of Telescopes, Apartado de correos 321, E-38700 Santa Cruz de la Palma, Canary Islands, Spain
	 \and
	     Now at Max Plank Institut {f\"ur Physik, F\"ohringer Ring 6, 80805 Muenchen, Germany}
	 \and
	     Now at Erlangen Centre for Astroparticle Physics, Erwin-Rommel-Str. 1, 91058 Erlangen, Germany
             }

   \date{Received ; accepted}

  \abstract
   {
    Very high-energy (VHE, E\,$>$\,100\,GeV) $\gamma$-ray data are a valuable input
    for multi-wavelength and
    multi-messenger (e.g. combination with neutrino
    data) studies.
   }
   {
    We aim at the conservation and homogenization of historical, current, and future {VHE $\gamma$-ray-data}
    on active galactic nuclei (AGN).
   }
   {
    We have collected lightcurve data taken by major VHE experiments since 1991
    and combined them into long-term lightcurves for several AGN, and now provide
    our collected datasets for further use.
    Due to the lack of common data formats in VHE $\gamma$-ray astronomy,
    we 
    have defined relevant datafields to be stored in standard data formats.
%chose to store data in the standard FITS file format.
    The time variability of the combined VHE lightcurve data was investigated,
    and correlation with archival X-ray data collected by {{\em RXTE}/ASM} tested.
   }
   {
    The combination of data on the prominent blazar Mrk\,421 
    from different experiments yields a lightcurve 
    spanning more than a decade.
    From this combined dataset we derive an integral baseline flux
    from Mrk\,421 that must be lower than 33\,\% of the Crab Nebula flux above 1\,TeV.
    The analysis of the time variability yields log-normal flux variations in the
    VHE-data on Mrk\,421.
   }
   {
    Existing VHE data contain valuable information concerning the variability of AGN and
    can be an important ingredient for multi-wavelength or multi-messenger studies.
    In the future, upcoming and planned experiments will provide more data from
    many transient objects, and the interaction of VHE astronomy with classical astronomy
    will intensify.
    In this context a unified and exchangeable data format
    will become increasingly important. {Our data collection is available at the url:
    {\tt {http://nuastro-zeuthen.desy.de/magic\_experiment/projects/light\_curve\_archive/index\_eng.html}}}.
   }

   \titlerunning{Long-term $\gamma$-ray lightcurves}
   \maketitle
%
%________________________________________________________________

\newcommand{\tevflux}{$\mathrm{F_{VHE}}$}
\newcommand{\asmflux}{$\mathrm{F_{ASM}}$}
\newcommand{\rhs}{$\mathrm{R_{HS}}$}
\newcommand{\thigh}{$\mathrm{T_{high}}$}
\newcommand{\tlow}{$\mathrm{T_{low}}$}

%\begin{abstract}
%% archive
%\end{abstract}

\section{Introduction}

%***\\
%
%OBSERVATORY CHARACTER OF CTA ...\\
%Write that an exchangable well defined data format is needed. We provide it...\\
%
%***\\

% AGN emission
The broad-band emission observed from active galactic nuclei (AGN) spans
the complete electromagnetic spectrum from radio to VHE (very high-energy, E\,$>$\,100\,GeV) $\gamma$-rays.
%
% AGN VHE discovery
Since the discovery of the blazar Mrk\,421~in the VHE regime \citep{1992Natur.358..477P},
many new detections of AGNs by different experiments have been reported
\citep[see, e.g.][for reviews]{2008RvMA...20..167H,2009ARA&A..47..523H}.
%
% hadronic / leptonic
%Even though a hadronic origin of the observed signals cannot be
%ruled out, leptonic acceleration mechanisms are favoured
%to explain the observed broad band emission.
%
% flux variations
Strong flux variations on different timescales were observed from many 
AGNs.
%
% variability
Variability in the VHE $\gamma$-ray regime was measured down to
the minute timescale \citep{2008PhLB..668..253M,2007ApJ...664L..71A}.
% lightcurves
Many AGN lightcurves were produced by different $\gamma$-ray
experiments from observations of flaring states, dedicated monitoring programs, or
from joint campaigns with other experiments.
Long-term $\gamma$-ray lightcurves are valuable to address open questions
concerning AGNs, such as possible periodicities \cite[see][an analysis using the data collection of the present
publication]{thieler:2010},
the log-normality of their flux state distribution, the nature of the radiation mechanism
of AGNs, or the estimation of their dutycycle.

A log-normal behavior can be indicative of a multiplicative
process of the underlying mechanism governing the variability of the object.
In the case of AGNs, this could be evidence for a connection of the observed emission to
accretion disk activity \citep[see][and references therein]{2009A&A...503..797G}.
Previously, log-normal flux variations were reported in the X-ray band from
two objects: the narrow-line Seyfert\,1 galaxy IRAS\,13244-3809 \citep{2004ApJ...612L..21G}
and BL\,Lac \citep{2009A&A...503..797G}. In VHE data, log-normal variability was observed
from a high flux state of the BL\,Lac object PKS\,2155-304 \citep{2008bves.confE..16D}.

The nature of the AGN radiation mechanism in the VHE regime,
i.e. whether the observed radiation has a leptonic or hadronic origin, still
remains ambiguous.
%The observed short-term variability is
%often used as an argument for a leptonic origin of the observed radiation.
%However, hadronic acceleration mechanisms or a hadronic flux component cannot be completely
%ruled out.
%
% MM: need high-state rate as input
The detection of neutrinos from these objects would 
prove the existence of a hadronic component.
% contribution to the observed VHE-signals.
In the framework of multi-messenger strategies 
(e.g., combining electromagnetic with neutrino data),
the phenomenology of lightcurves in the electromagnetic
wavelength band can give valuable input. 
The Neutrino triggered Target of Opportunity (NToO) program \citep{palaiseau2005eb,2008ICRC....3.1257A}
is based on the idea of
neutrino events from a variable object (single events or multiplets) that are used to trigger
VHE monitoring of the same object. Coincidences between neutrino triggers and 
$\gamma$-ray flux high states, which occur more often than expected
from random coincidence with atmospheric neutrinos,
would be evidence of a hadronic component
of the VHE $\gamma$-ray signal and a cosmic origin of the neutrinos.
Long-term $\gamma$-ray lightcurves can be used to estimate the AGN dutycycle,
which is an important input parameter for such analyses.
\section{Lightcurve data}
Considering the heterogeneous nature of historical and present VHE data
comprising different file and content formats from different experiments,
a common data format is desirable.
Moreover, as described above, VHE astronomy has already started to interface strongly with different fields of
classical astronomy. In view of an effective exchange and diffusion of VHE data within the astronomy
and astroparticle
community, our storage strategy is to use an ascii file format (SLF) and the widely used FITS and VOTable
file formats (see following section).

The use of a standard data format will also become increasingly important in the framework of the 
next-generation Cherenkov telescope systems, such as the upcoming CTA (Cherenkov Telescope Array)
experiment \citep[see e.g.][]{2008AIPC.1085..824M}. In the CTA era, VHE astronomy
will intensify multiwavelength (and multimessenger) interactions with other fields.
Furthermore, a standard data format is essential for running an experiment
like an observatory, and make it open to the whole astronomy community.

Publicly available lightcurve data from 1992 until today were
collected from the
Whipple \citep{1995ApJ...438L..59K,1996ApJ...460..644S,1996ApJ...472L...9B,1999ApJ...526L..81M},
HEGRA \citep{1999A&A...342...69A,1999A&A...349...29A,2001ApJ...559..187K,2001ApJ...546..898A,2002A&A...393...89A,2003A&A...410..813A,2004ApJ...614..897A,kestel},
CAT \citep{2000PhDT.........6P,2001A&A...374..895P},
H.E.S.S. \citep{2005A&A...437...95A,2006A&A...457..899A},
MAGIC \citep{2008ApJ...674.1037A,2009ApJ...691L..13D},
and VERITAS \citep{2006ApJ...641..740R,2009ApJ...691L..13D} experiments.
% only lightcurves
%Information on the energy spectrum
%in different flux states is only partly available.
%
{We are also working on collecting more data from these and other experiments such as the
Crimean GT-48 Cherenkov telescope \citep[e.g.][]{2007BCrAO.103...16N},
the Tibet air-shower array \citep[e.g.][]{2003ApJ...598..242A},
or the Patchmarhi Cherenkov Telescope Array \citep[e.g.][]{2008ChJAA...8..395G}}.
\subsection{Data collection and formatting}
% formats
Different formats and standards are used by different experiments. 
{We use simple directly usable ascii tables and the
standard astronomy file formats FITS and VOTable} to store the collected, combined data.
In the present work, the types and units of the FITS/VOTable datafields (i.e. columns) follow
a simple lightcurve format (hereafter SLF) given in Table~\ref{slfformat}.
The data are also provided in ascii file form following the SLF definition. These files are
referred to as slf-files.
The tables contain rows of (typically nightwise) integral flux-measurements
and further information necessary for subsequent analyses.
{The units of the datafields are defined within the FITS/VOTable files and follow
the conventions defined in Table~\ref{slfformat} within the slf-files (ascii).}
%In addition to the FITS format files, we provide simple ascii tables that contain the same data
%in the simple lightcurve format, SLF.
\begin{center}
\begin{table*}[ht]
\centering
 \begin{tabular}{lllll}\hline
 \bf Datafield	&	\bf Name		& \bf TTYPE		& \bf TUNIT&  \bf Explanation	\\\hline
 1		&	$MJD_{mid}$		& mjd\_mid\_exp		& MJD		& center of observation exposure \\
 2 		&	$MJD_{start}$		& mjd\_start		& MJD		& start of observation \\
 3      	&	$MJD_{end}$		& mjd\_end		& MJD		& end of observation \\
 4      	&	$F$ 			& int\_flux		& Crab units 	& Observed integral flux\\
 5      	&	$\Delta F_{stat}$	& sigma\_int\_flux\_stat& Crab units 	& Statistical error on $F$\\
 6      	&	$\Delta F_{sys}$	& sigma\_int\_flux\_sys	& Crab units 	&Systematical error on $F$\\
 7      	&	$\alpha$		& alpha			& none 		& Differential spectral index\\
 8      	&	$\Delta\alpha_{stat}$	& sigma\_alpha\_stat	& none 		& Statistical error on $\alpha$\\
 9      	&	$\Delta\alpha_{sys}$	& sigma\_alpha\_sys	& none 		& Systematical error on $\alpha$\\
 10     	&	$E_{thr}$		& e\_thr		& TeV 		& Energy threshold in TeV\\
 11		&	$E_{cutoff}$		& e\_cut		& TeV 		& Cutoff energy in TeV\\
 12		&	Experiment		& experiment		& none 		& Experiment string\\
 13		&	Duration		& duration		& hours 	& Duration of observation\\ 
 14		&	Reference		& reference		& none		& Reference string (pref. ADS format)\\
 15		&	Flux-flag		& fflag			& none		& flux-measurement: '$=$', upper limit: '$<$'\\
 16		&	additional entry 1	& add\_entry1		& 		& e.g., special parametrizations\\
 17		&	additional entry 2	& add\_entry2		& 		&	\\
 ...		&	...			& ...			& 		&	\\\hline
 \end{tabular}
 \caption{\label{slfformat}Datafield definition$^*$ used for storage of lightcurve data in the FITS/VOTable file format.
	\newline
 {\footnotesize $^*$ The used datafield definition is referred to as simple lightcurve format (SLF) in the text.
 TTYPE and TUNIT are the name and units used within the FITS file.}
 }
\end{table*}
\end{center}
The SLF datafield format definition provides the means to effectively combine heterogeneous datasets.
 In case further information than included in columns 1 to 12 are necessary, additional entries are foreseen.
 For example, to define a spectral shape different from the standard
 pure powerlaw, the spectral parameterization together with the additional parameters must be defined as comments
 within the data files.
In case of lightcurve data only containing information on a few datafields, the flexibility
of the binary FITS and VOTable file format allows minimizing the data volume.
For example, if only the observation date and the flux value are known for the
total lightcurve data of an object, the binary file can only contain $MJD_{start}$ and $F$,
along with the definition of units used.
While this example represents the simplest case of a homogeneous dataset,
in practice, the heterogeneous lightcurve data combinations from different
experiments can include information on a specific datafield for one subset of
the combined data (i.e. for one experiment), and not for another subset.
In this (most common) case, when information is not available for datafields of
some of the measurements, the following conventions for default datafield values apply.
\begin{itemize}
% \item Comment lines have to start with the symbol \#.
 \item If no start and end times are known, {then $MJD_{start}$ is set to the same value as
        $MJD_{mid}$, i.e. the middle of the exposure time,} and $MJD_{end}$ to -1. In this case the duration might still
	be given.
 \item {In some cases only the MJD of the observations day is known. Then, $MJD_{mid}$ is the MJD of the observation day.}
 \item Most energy spectra can be described by a pure powerlaw 
       (differential spectral index $\alpha$) or a powerlaw with an exponential
       cutoff (cutoff energy $E_{cutoff}$). 
       Additional fields are foreseen for cases with different parameterizations
       of the energy spectrum. If such a different parameterization is given, a defining
       formula containing the relevant fields has to be written in the comment section of the lightcurve file.
 \item {The flux-flag (fflag) is a one-character flag indicating whether the integral flux $F$ is an upper limit
       (fflag = '$<$') or a flux measurement (fflag = '=').}
 \item In the present work, we chose Crab Nebula flux units for the observed integral flux
       $F$ above the energy threshold $E_{thr}$.
 \item For any other entry, the value -1 means that no information is available.
\end{itemize}
\subsection{Lightcurve combination}
% combination
For a combination of lightcurves from different experiments, the measured 
integral flux values must be converted to a common energy threshold.
For the results in this paper, we chose an energy threshold of 1\,TeV.
The conversion of integral fluxes in Crab units to the same energy threshold
requires knowledge of the energy spectrum of the considered object
and of the Crab Nebula as measured by the same experiment.
The differential energy spectrum 
of the Crab Nebula in the energy range covered by most 
VHE $\gamma$-ray experiments is described well by a pure powerlaw 
of the form $\phi_{Crab}(E) = \phi^{(0)}_{Crab}\cdot (E/TeV)^{-\Gamma_{Crab}}$
in the energy range from 100\,GeV to few tens of TeV.
Deviations from the pure powerlaw form at both ends of this
energy range
(as expected in the framework of the inverse Compton scenario) were not
taken into account here. Extrapolation to energies below 100\,GeV
requires a full parameterization of the spectral energy 
distribution of the Crab Nebula \cite[e.g.][]{2004ApJ...614..897A}.
An integral flux in Crab units above the energy threshold $E_{thr}$ is given by
% equation for combination
\begin{equation}
\label{eqcombination0}
F(E>E_{thr}) = \frac{\int_{E_{thr}}^\infty{dE \phi(E)}}{\int_{E_{thr}}^\infty{dE \phi_{Crab}(E)}} 
%= \frac{\int_{E_{thr}}^\infty{dE \phi(E)}}{\phi^{(0)}_{Crab} \cdot E_{thr}^{-\Gamma_{Crab}+1}} 
= \frac{\int_{E_{thr}}^\infty{dE \phi(E)}}{F_{Crab}(E>E_{thr})}
\end{equation}
where $\phi(E)$ and $\phi_{Crab}(E)$ are the differential energy spectra of the source
and of the Crab Nebula (in units of photons\,TeV$^{-1}$cm$^{-2}$s$^{-1}$).
The corresponding integral fluxes above a given energy threshold are denoted
with $F(E>E_{thr})$.
%In case of known differential energy spectrum $\phi(E)$ of any form,
%Equation~\ref{eqcombination0} can be evaluated numerically.
%In many observations, the considered VHE $\gamma$-ray source yields
In the case of a pure powerlaw differential energy spectrum $\phi(E)$,
the conversion of an observed integral flux $F(E>E_0)$ to a given
threshold energy $E_{thr}$ is given by
\begin{equation}
\label{eqcombination}
F(E>E_{thr}) = \left(\frac{E_{thr}}{E_{0}}\right)^{-\Gamma+1} \frac{F(E>E_{0})}{F_{Crab}(E>E_{thr})},
\end{equation}
where $E_0$ is the energy threshold of the observation.
% numerical integration in case of deviating spectra

\begin{figure*}[ht]
\centering
 \includegraphics[width=\textwidth]{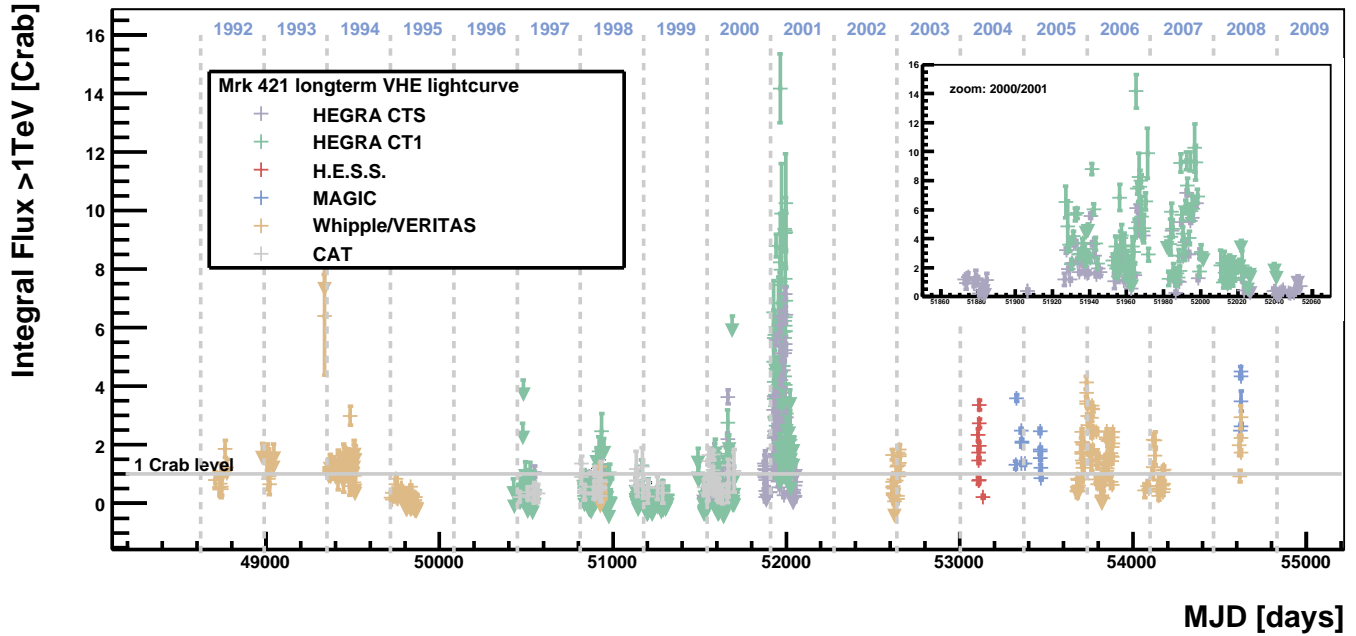}
 \caption{\label{mrk421lc} Long-term lightcurve of Mrk\,421 (day-wise integral flux). Data from the major 
	  $\gamma$-ray telescopes were combined and normalized to the same energy
	  threshold (1\,TeV) and converted to Crab units (see text). A zoom into the period of strong
	  activity (2000/2001) is also shown.}
\end{figure*}

% table for Crab Spectra 
% results Mrk421
The combined day-wise, integral flux lightcurve of the BL\,Lac object Mrk\,421 above 1\,TeV
is shown in Figure~\ref{mrk421lc}. This lightcurve includes all data we have collected
so far, covering an uprecedented 17-year time-span from 1992 to 2008.
\begin{figure*}[ht]
\centering
 \includegraphics[width=\textwidth]{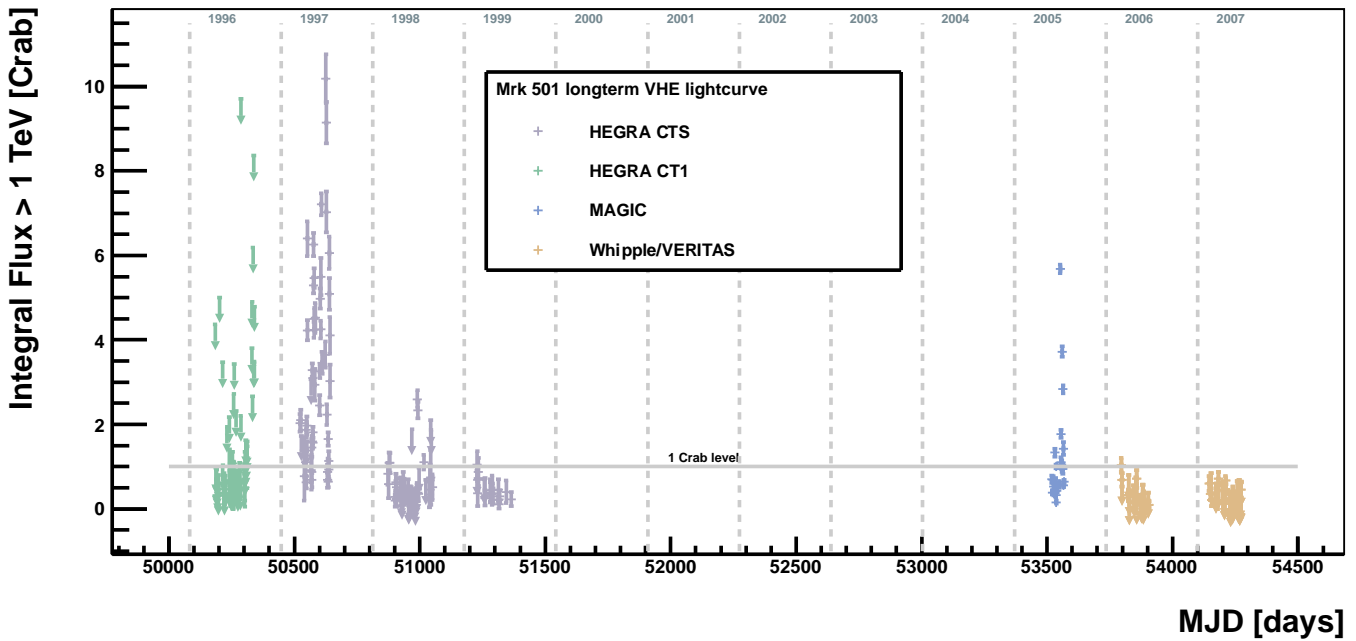}
 \caption{\label{mrk501lc} Long-term lightcurve of Mrk\,501. Available data
	  were combined and normalized. Shown are the day-wise integral fluxes above 1\,TeV
          in units of the Crab Nebula flux.}
\end{figure*}
In Figure~\ref{mrk501lc}, the day-wise integral flux lightcurve of the BL\,Lac object
Mrk\,501 is shown.

\subsection{Systematic errors}
% generic / intro
The inhomogeneity of data combined from different experiments and the
inclusion of data from the pioneering time of VHE $\gamma$-ray astronomy
induces systematic errors that are not easily evaluated.
Combination of data from many different experiments,
different time periods, partly inaccurate flux normalizations (i.e. $\gamma$-ray rate measurements) and energy thresholds 
can lead to systematic errors in the relative flux normalizations.
When combining data, the flux normalization depends on assumptions on
the spectral shape and the energy threshold.
Here, limiting factors can be experimental uncertainties (especially in old data)
induced by poor knowledge of atmospheric conditions,
aging detector effects, and intrinsic spectral variability of the considered objects.
% Use Crab Nebula data from same experiment reduces systematic error
We convert flux and rate measurements to Crab Nebula flux units using
Crab Nebula data taken by the same experiment as close in time as possible,
therewith reducing systematic uncertainties induced by seasonal or instrumental
variations.

% Flux normalization depends on measured energy spectrum / chosen energy threshold.
% Combination strongly depends on spectral shape.
%Spectral reconstruction can be critical in different ways.
%While a parameterization 
%by a powerlaw or a powerlaw with an exponential cutoff is often the best choice to fit
%data from one exeriment, the overall spectral energy distribution in the framework
%of inverse Compton models is better described by a log-parabola. In case of a 
%powerlaw with an exponential cutoff, the hardness of the spectral powerlaw index
%can be overestimated. Therefore, data conversion to lower energy thresholds
%tends to yield underestimated flux levels.
%Furthermore, even if two experiments with the same energy threshold
%might yield identical flux levels from simultaneous observations, a small difference in
%reconstructed spectral shape might lead to deviations when converting to different thresholds. 
The error on the reconstructed spectral shape results in an error on the flux state
when converting to different thresholds. 
For example, when converting integral fluxes from 1\,TeV to 100\,GeV, typical systematic errors on the spectral index of 0.1
(pure powerlaw case) lead to a relative systematic error on the integral flux of 30\,\% due to this extrapolation.
%(see Table~\ref{errors}).
The datafields contained in our FITS files provide all information needed to estimate these systematic errors
for conversion to different energy thresholds.
A conversion over a narrower energy range than used in this example limits this systematic error to
less than 30\,\%.
%\begin{table}{ht}
%\begin{tabular}{lr}
%\end{tabular}
%\end{table}

%
The flux units used in some publications are given in uncalibrated units
(e.g. counts per minute), i.e. as an instrument-specific $\gamma$-ray count-rate. 
% unknown energy threshold 
The energy threshold is also unknown in some cases.
%This is true especially for older data.
For flux calibration of lightcurves from the Whipple experiment
before 1998, we used Crab Nebula data taken by the same experiment as close
in time as possible
\citep[see][and references therein]{1998ApJ...503..744H}.
%\citep{2008ICRC....2..691G,hegra???,hess???,cat???,???}.
However, due to dependencies on zenith angle and weather conditions,
normalization of rate measurements is very uncertain.
We estimate that the systematic error induced when including rate measurements
is below 40\,\%.
This form of relative systematic error can be avoided when restricting analysis
to flux measurements, by restricting to data from a single
experiment,
or by intercalibrating the different measurements when observations from different experiments overlap.

% gaps
%In VHE $\gamma$-ray astronomy, observations are restricted to night-time for most experiments.
%Additionally, an object is only visible to any ground-based experiment for a period of 
%a few months per year. The resulting irregular gaps in the lightcurves are naturally 
%unavoidable
%and have to be taken into account in any time-variability analysis.
%and represent a further limiting factor that has to be kept in mind
%for any analysis aiming at extracting variability information from these data.

%The above limitations and systematic uncertainties
%restrict any high-precision variability analysis to datasets from one experiment or
%to contemporaneous observations from different experiments that
%can be inter-calibrated.
%A homogeneous well-calibrated dataset would be preferable,
%nevertheless, all necessary information for an estimation of the systematic uncertainties
%is included in our data collection.
%Current standards and the quality of existing and future experiments
%will yield higher-quality datasets.
%Even today, the majority of data collected in this work are well calibrated data.
%Ongoing and future monitoring campaigns will contribute to a better
%time-coverage of interesting objects.

%\section{Correlation of VHE $\gamma$-rays with X-rays}
\section{Discussion}
{Significant evidence of a correlated gamma- and X-ray emission of blazars was presented
in earlier studies \citep[see e.g.][]{1996ApJ...470L..89T,1996ApJ...472L...9B,1999APh....11..189M,2001AIPC..599..694K,2007BCrAO.103...16N}.
Usually the reported correlation is linear, but in
a few cases a quadratic relation between the fluxes in both bands was found
\citep[see e.g.][]{2000A&A...353...97K}.
Such a correlation provides us with essential information on the
underlying acceleration and emission processes and is especially valuable for variability
studies. Very often the gamma- X-ray correlation is interpreted as a strong argument in
favor of the so-called synchrotron-Compton jet emission models in which the same population
of ultrarelativistic electrons is responsible for production of both X-rays
via synchrotron radiation and TeV $\gamma$-rays via inverse Compton scattering \citep{2005A&A...433..479K}.
However, it can also be accommodated in the hadronic framework in particular in models
which assume that the observed $\gamma$-ray emission is a result of interactions of accelerated
protons and ambient gas or low-frequency radiation \citep{2000NewA....5..377A,2003APh....18..593M}.}

Using contemporanous X-ray data extracted from the {\em RXTE}/ASM database web interface at 
MIT\footnote{\it http://xte.mit.edu/ASM\_lc.html}, we calculated correlation coefficients
between the VHE and X-ray bands.
For each VHE measurement, an average ASM count rate was calculated centered on the VHE observation date ($\pm$\,6 hours).
Only using measurements with a significance of at least 3 standard deviations for the VHE
data and 10 counts for the ASM measurement, these data yield correlation coefficients {(Spearman rank)} of
0.65 for Mrk\,421
%Pearsons correlation coefficient =  0.559799736308  p-value =  1.84944219098e-27
%Spearman correlation coefficient =  0.648692724704  p-value =  4.00900533896e-39
%correlation coefficient = 0.559800
and 0.68 for Mrk\,501.
%Pearsons correlation coefficient =  0.693844876147  p-value =  8.39687264694e-09
%Spearman correlation coefficient =  0.684849217868  p-value =  1.55279106423e-08
%correlation coefficient = 0.693845
{
In both cases, the probability of an uncorrelated system producing datasets with similar Spearman rank correlation coefficients
is less than $10^{-8}$.
}
The flux measurements in both wavelength bands are not strictly simultaneous.
However, they represent a measure for the average daily flux state of the objects,
and the observed behavior is consistent with correlated average VHE/X-ray daily flux levels modulated
by shorter term flux variations.

\begin{figure*}[Ht!]
 \parbox{0.49\textwidth}{\includegraphics[width=0.49\textwidth]{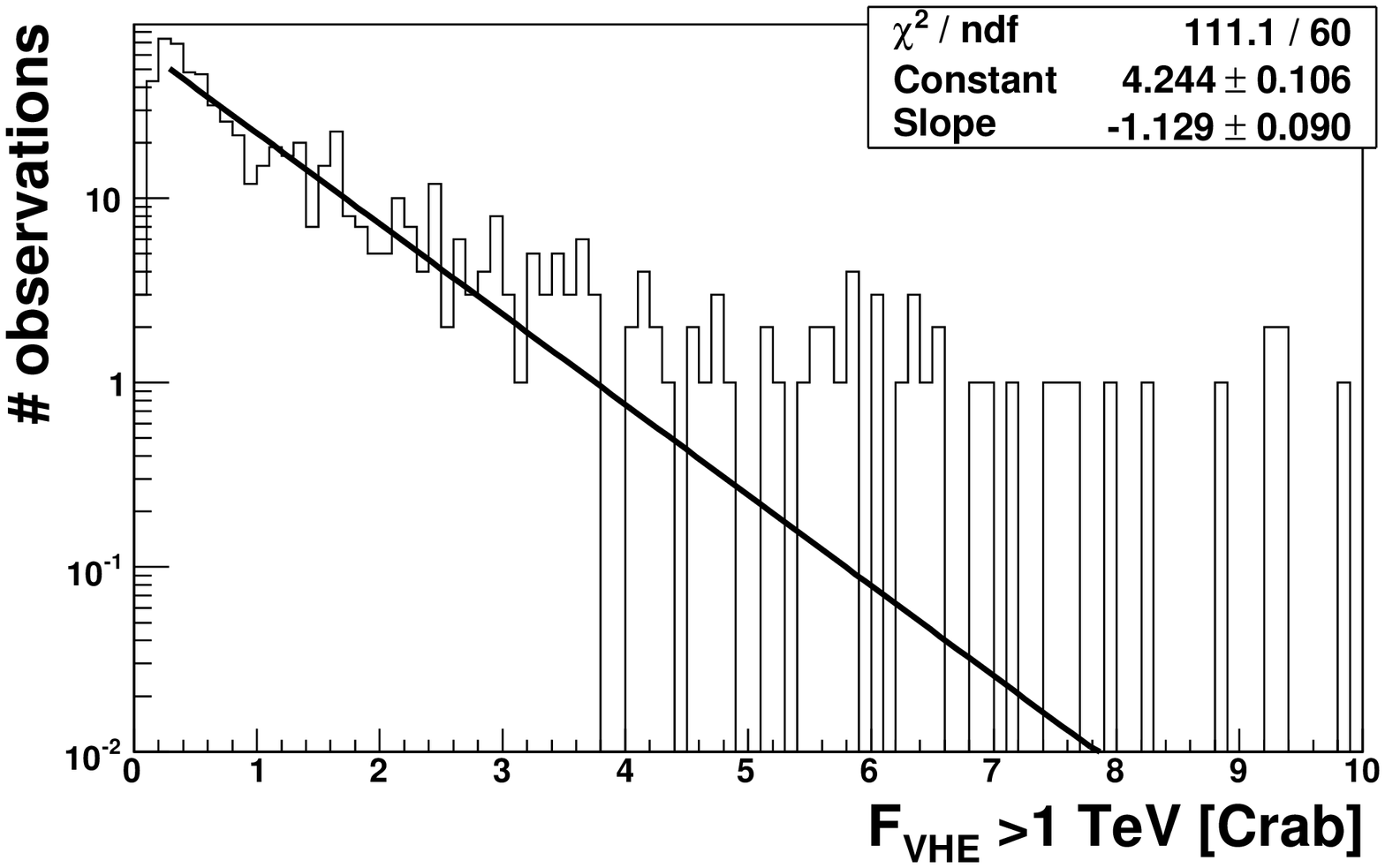}}\hfill
 \parbox{0.49\textwidth}{\includegraphics[width=0.49\textwidth]{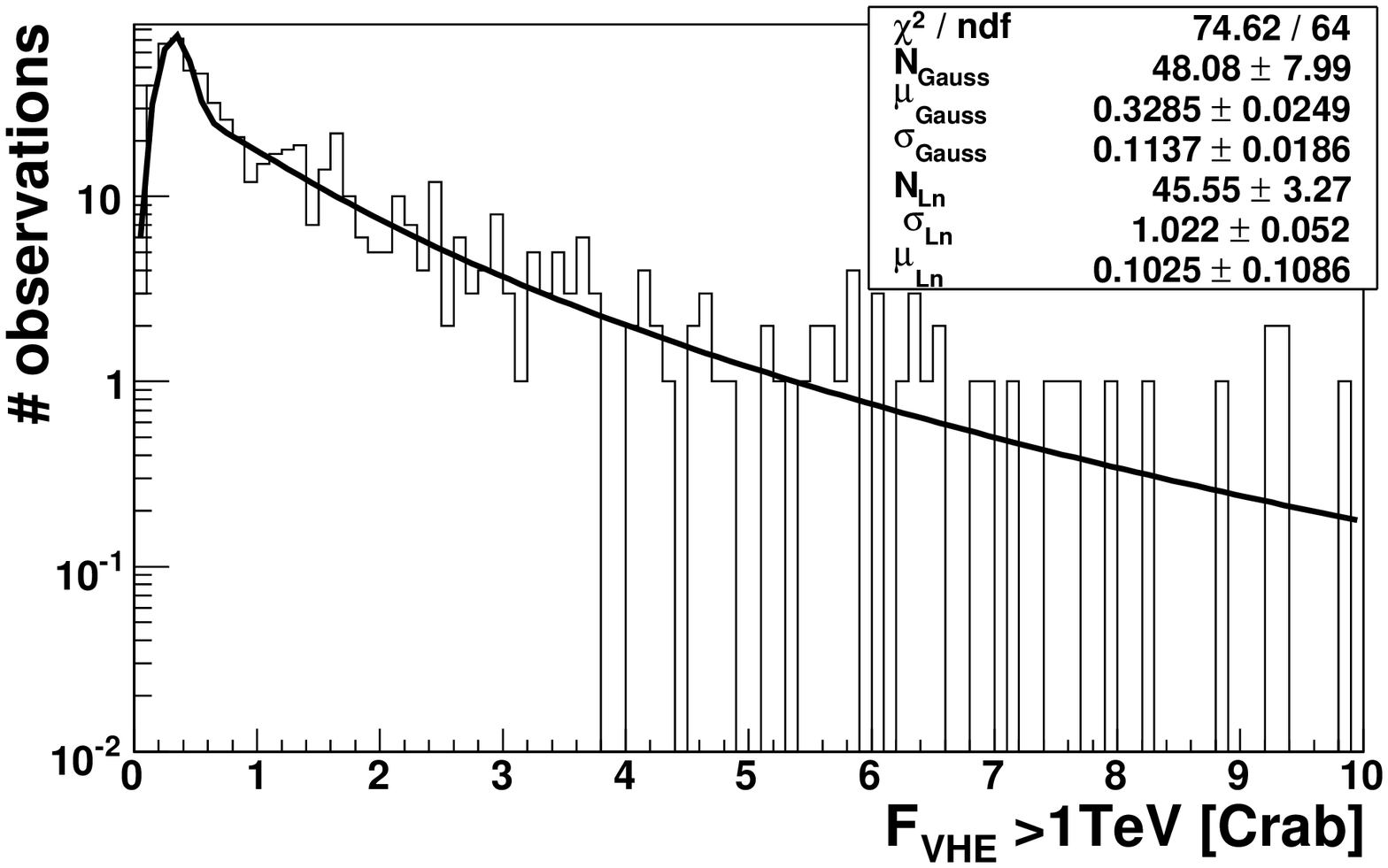}}
 \caption{\label{fluxstates} Distribution of VHE flux states
          of Mrk\,421. An exponential function fit above a flux of 0.25\,Crab,
          (avoiding detector threshold effects) can describe the data (left).
	  %Deviations from the exponential form at higher fluxes might be partly
	  %induced by observational biases (see text).
	  The data are very well described by a fit of a Gaussian+log-normal distribution (right).}
\end{figure*}

\begin{figure}[ht]
\centering
 \includegraphics[width=0.5\textwidth]{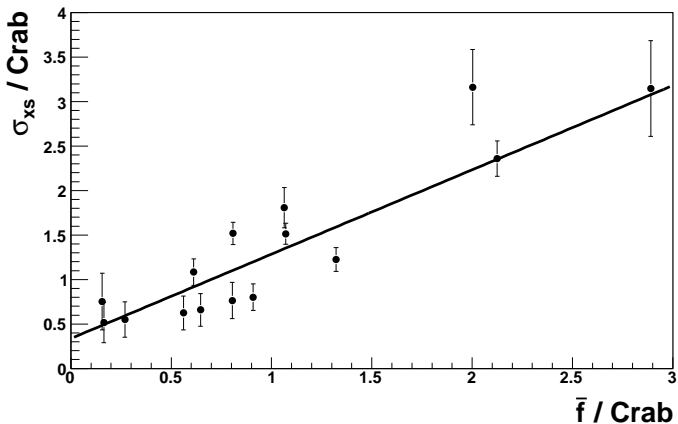}
 \caption{\label{sigmaxs}The excess variance $\sigma_{xs}$ as a function of the average flux within equal-length
 intervals. A line fit corresponding to $\sigma_{xs} \thicksim (0.95\,\pm\,0.10)\,f_i$ is also given.}
\end{figure}

%\section{Flux-State Distributions}
In Figure~\ref{fluxstates}, the distribution of the VHE flux states of Mrk\,421
(also see Figure~\ref{mrk421lc}) is shown. {Each entry of this histogram represents
a snapshot flux measurement of Mrk\,421 as given by the combined dataset.}
% how did I produce this distribution ?
All individual observed integral flux values were converted to flux values in units of the 
Crab Nebula flux and normalized to a common energy threshold of 1\,TeV.
The observed energy spectra of the Crab Nebula and Mrk\,421 as observed
by the individual experiments were taken into account.
% The resulting integral flux values were weighted with their observation time.
% exponential fit
{The overall distribution can be described by an exponential law (left panel) above a flux of a few tenths of Crab.
An exponential shape might indicate that the measured flux states mainly
reflect a stochastic outburst state of the object.
%
%At higher flux values, {deviations from the exponential shape are seen}.
%{Here,} the distribution {might be} biased due to external-triggering
%(e.g. joint observations, multi-wavelength campaigns) 
%and self-triggering on high flux states (e.g. expansion of 
%observation window during monitoring campaign).
%However, it is not possible to quantify this bias, and its effect
%on the observed flux state distribution might well be negligible.
%
{As shown in the right panel of Figure~\ref{fluxstates}}, a better fit to the data is obtained when
using the sum of a Gaussian and a log-normal distribution
\citep{aitchison:1957,limpert:2001},
as given by
\begin{equation}
  f(x) = \frac{N_{Ln}}{x \sigma \sqrt{2\pi}} exp\left(-\frac{(log(x)-\mu)^2}{2 \sigma^2}\right).
\end{equation}
% sigma_xs
The data were further divided into time intervals of equal length, each comprising 40 flux observations $f_i$ with
statistical errors $\sigma_i$.
The excess variance $\sigma_{xs} = \sqrt{\frac{1}{N}\sum^{N}_{i=1}(f_i - \bar{f})^2 - \sigma_i^2}$
\citep{2003MNRAS.345.1271V} was calculated for each time interval.
The excess variance $\sigma_{xs}$ is a measure of the Poisson noise corrected rms value of the fluxes within the
corresponding time interval.
%\begin{figure*}[ht]
% \parbox{0.49\textwidth}{\includegraphics[width=0.49\textwidth]{Mrk421DailyFluxstatesTeV.eps}}
% \parbox{0.49\textwidth}{\includegraphics[width=0.49\textwidth]{Mrk421FluxStateComponents.eps}}
% \caption{\label{fluxstates} \emph{Left:} Distribution of VHE flux states
%          of Mrk\,421. An exponential function was fitted above a flux of 0.5\,Crab,
%          avoiding detector threshold effects.
%	  \emph{Right:} The distribution of flux states of Mrk421 as measured by the ASM can
%          be interpreted as a baseline (medium grey), a stochastic outburst distribution 
%	  (dark grey) and a very low level of detector noise (light grey).}
%\end{figure*}
%
Figure~\ref{sigmaxs} shows $\sigma_{xs}$ as a function of the average flux $\bar{f}$ within the interval.
A clear proportionality is seen with $\sigma_{xs} \thicksim (0.95\,\pm\,0.10)\,f_i$.
A log-normally shaped distribution with a proportionality of the excess variance
with the flux are evidence of log-normal flux variations \citep{aitchison:1957}.

Below a flux level of few tenths of Crab, one might expect a low baseline flux level
that can be described by a Gaussian.
At these low fluxes, however, detector sensitivity threshold effects become 
increasingly} important, i.e. lower fluxes cannot be detected significantly within the short observation
windows.
Therefore, the mean of the Gaussian
fit (0.33 Crab flux units) must be treated as an absolute upper limit on the integral baseline flux
above 1\,TeV from Mrk\,421.
%The distribution of ASM count rates (Figure~\ref{fluxstates}, right panel)
%can be interpreted using 3 components. A population of low count rates might be interpreted as 
%a baseline flux (dashed line, Gaussian).
%A stochastic outburst distribution is described by an exponential function (solid line).
%An additional Gaussian (dotted line) describes a very low level of detector noise
%that does not affect the overall distribution at higher fluxes.
%One has to note that the observed distribution is the true distribution folded
%with the detector acceptance. Therefore, at low count rates the distribution is strongly
%dominated by systematic effects (detector sensitivity threshold).
%The true baseline -- if existent at all -- might be lower.
%
More information such as the variability type (e.g. red noise / white noise / blue noise)
or the duty-cycle of the object can be extracted from
the flux state distributions. Previously, first steps in this direction
were taken by \citet[][]{tluczykont:2006a};
however, further investigations are necessary using simulations and
will be the topic of a subsequent publication.

With increasing exposure achieved by the major current generation experiments more
data will soon be available. As discussed previously, one important aspect for long-term
variability studies will be to carry out unbiased (random) observations
to avoid systematic selection effects.

\section{Summary \& outlook}
% most long-term ever lightcurve of Mrk421 !
VHE $\gamma$-ray flux measurements of AGN have been collected from observations of
different experiments since 1992.
For the first time these data were combined into single long-term lightcurves,
and are provided for further analysis in the standard FITS file format.
%Alternativley an ascii-based simple lightcurve format (SLF) is also provided.
The BL Lac object Mrk\,421 yields the most extended dataset with a combined
lightcurve spanning more than a decade.
The collected data is publicly available for download\footnote{\tt \scriptsize http://nuastro-zeuthen.desy.de/magic\_experiment/\newline
projects/light\_curve\_archive/index\_eng.html}
%{\tt \tiny http://nuastro-zeuthen.desy.de/magic\_experiment/projects/light\_curve\_archive/index\_eng.html}

% correlation
The observed flux states {averaged over 12\,h} of Mrk\,421 and Mrk\,501 above
1\,TeV are consistent within statistical and systematic errors with a linear correlation
with daily averaged \emph{RXTE}/ASM X-ray count rates, with correlation coefficients of {0.65 and 0.68} respectively.
%0.65 in the case of Mrk\,421
%Pearsons correlation coefficient =  0.559799736308  p-value =  1.84944219098e-27
%Spearman correlation coefficient =  0.648692724704  p-value =  4.00900533896e-39
%correlation coefficient = 0.559800
%and 0.68 in the case of Mrk\,501.
%Pearsons correlation coefficient =  0.693844876147  p-value =  8.39687264694e-09
%Spearman correlation coefficient =  0.684849217868  p-value =  1.55279106423e-08
%correlation coefficient = 0.693845

% flux state distribution
The long-term flux state distribution of Mrk\,421 can be well described by the
sum of a Gaussian and a log-normal distribution. This behavior can be interpreted
by the observation of a baseline flux and
a stochastic flux state population governed by an underlying multiplicative process.
% baseline flux
In the framework of this interpretation,
the combined dataset on Mrk\,421 allowed setting an absolute upper limit on the
baseline flux of Mrk\,421 of 0.33 Crab flux units above 1\,TeV.

Unbiased data from monitoring campaigns by ongoing
experiments \citep[for a review see][]{2008AIPC.1085....3W}
%\citep[H.E.S.S., MAGIC, VERITAS][]{hess,magic,veritas}
and data from future experiments such as CTA will further extend the
long-term coverage of variable objects.
A unified standard data format will become increasingly important
in the future, when more high-sensitivity VHE data becomes available,
and when interfacing with other fields of classical astronomy and astroparticle physics
(multi-wavelength and multi-messenger) intensifies.
We intend to store these future data as described in this work
in standard data formats using the introduced SLF field structure definition.
%

%Monitoring campaigns of transient VHE $\gamma$-ray sources, a higher data transparency
%as well as unified data standards are necessary.
%Future data from monitoring campaigns of transient VHE $\gamma$-ray sources
%will provide rich data that we intend to archive in slf format

\begin{acknowledgements}
We thank all colleagues who provided data in electronic form to us.
MT thanks Dr. B. Giebels and Prof. Dr. D. Horns for valuable discussions.
We thank Dr. B. Pence for helpful comments on an ealier draft.
This research has made
use of NASA's Astrophysics Data System Bibliographic Services.
\end{acknowledgements}

\bibliography{tluczykont_lc}

\end{document}